\documentclass{revtex4}
\usepackage{amsfonts}
\usepackage{amsmath,amssymb}
\begin{document}

\title{Aspects of Diffeomorphism Invariant Theory of Extended Objects}
\author{V.~G. Gueorguiev
\footnote{
\uppercase{O}n leave of absence from 
\uppercase{I}nstitute of \uppercase{N}uclear \uppercase{R}esearch and \uppercase{N}uclear \uppercase{E}nergy, \uppercase{B}ulgarian \uppercase{A}cademy of \uppercase{S}ciences, \uppercase{S}ofia 1784, \uppercase{B}ulgaria.}
\footnote{
\uppercase{R}esearch in part supported by the \uppercase{U.S.} \uppercase{N}ational \uppercase{S}cience \uppercase{F}oundation under \uppercase{G}rant \uppercase{N}o. \uppercase{PHY} 0140300, the {\it 3rd \uppercase{I}nternational \uppercase{S}ymposium on \uppercase{Q}uantum \uppercase{T}heory and \uppercase{S}ymmetries}, and the {\it \uppercase{A}rgonne \uppercase{W}orkshop on \uppercase{B}ranes and \uppercase{G}eneralized \uppercase{D}ynamics}.}}
\address{Department of Physics and Astronomy, \\ Louisiana State University, Baton Rouge, LA 70803 \\ E-mail: vesselin@phys.lsu.edu }

\begin{abstract}
The structure of a diffeomorphism invariant Lagrangians for an extended object W embedded in a bulk space M is discussed by following a close analogy with the relativistic particle in electromagnetic field as a system that is reparametrization-invariant. The current construction naturally contains, relativistic point particle, string theory, and Dirac--Nambu--Goto Lagrangians with Wess--Zumino terms.  For ÒLorentzian metricÓ field, the non-relativistic theory of an integrally submerged W-brane is well defined provided that the brane does not alter the background interaction fields. A natural time gauge is fixed by the integral submergence (sub-manifold structure) within a Lorentzian signature structure. A generally covariant relativistic theory for the discussed brane Lagrangians is also discussed. The mass-shell constraint and the Klein--Gordon equation are shown to be universal when gravity-like interaction is present. A construction of the Dirac equation for the W-brane that circumvents some of the problems associated with diffeomorphism invariance of such Lagrangians by promoting the velocity coordinates into a non-commuting gamma variables is presented.
\end{abstract}
\maketitle

\textbf{Introduction.} The Hamiltonian and Lagrangian formulation\cite{Kilmister 1967,Goldstain 1980} are two very useful approaches in physics. In general, these two approaches are related by the Legander transformation. For a reparametrization-invariant theory, however, there are problems in changing from the Lagrangian to the Hamiltonian approach.\cite{Goldstain 1980,Gracia and Josep,Rund 1966,Lanczos 1970} In this paper the focus is on the properties of reparametrization-invariant matter systems such as, the relativistic particle and its extended object (brane) generalization within the Lagrangian approach. We try to answer the question: ``What is the Lagrangian for an extended `matter' object?''

\textbf{Matter Lagrangian for relativistic particle.} The action for a massive relativistic particle has a nice geometrical meaning: it is the distance along the particle trajectory\cite{Pauli 1958} provided that the units are such that $x^{0}=ct $ and the particle moves with a constant 4-velocity ($g_{\mu \nu}v^{\mu}v^{\nu}=1$): 
\begin{equation} \label{S1} 
S_{1}=\int d\tau L_{1}(x,v)=\int d\tau\sqrt{g_{\mu \nu}v^{\mu}v^{\nu}}
\rightarrow \int d\tau.
\end{equation} 
For a massless particle, such as a photon, the length of the 4-velocity is zero ($ g_{\mu \nu}v^{\mu}v^{\nu}=0 $) and the appropriate `good' action\cite{Pauli 1958} is: 
\begin{equation}\label{S2} S_{2}=\int L_{2}(x,v)d\tau =\int g_{\mu
\nu}v^{\mu}v^{\nu}d\tau.
\end{equation} 
The Euler--Lagrange equations obtained from $S_{1} $ and $S_{2}$ are equivalent, even more, they are equivalent to the geodesic equation as well: 
\begin{equation}\label{geodesic equations}
\frac{d}{d\tau}\vec{v}=D_{\vec{v}}\vec{v}=v^{\beta}\nabla_{\beta}\vec{v}=0
\end{equation}
Since the Levi--Civita connection $ \nabla $ preserves the length of the vectors\cite{Pauli 1958} ($ \nabla g(\vec{v},\vec{v})=0 $) this equivalence is not surprising because the Lagrangians in  (\ref{S1}) and (\ref{S2}) are functions of the preserved arc length $g(\vec{v},\vec{v})=\vec{v}^{2} $. However, the equivalence between $S_{1}$ and $S_{2}$ has a much deeper roots. 

\textbf{Homogeneous Lagrangians.} 
Since $L_{2}$ is a homogeneous function of order $2$ with respect to $\vec{v}$, the corresponding Hamiltonian function ($ h=v\partial L/\partial v-L $) is exactly equal to $L_{2}$. Thus $ L_{2} $ is conserved, and so is the length of $ \vec{v} $. Any homogeneous Lagrangian in $ \vec{v} $ of order $ n\neq 1 $ is conserved because $ h=(n-1)L $. If $ dL/d\tau =0 $, then the Euler--Lagrange equations for $ L $ and $ \tilde{L}=f\left(L\right) $ are equivalent. This is an equivalence that applies to homogeneous Lagrangians in particular. It is different from the usual equivalence $ L\rightarrow \tilde{L}=L+d\Lambda /d\tau $ or the more general equivalence discussed by Hojman and Harleston Ref.~\cite{Hojman and Harleston}. Any solution of the Euler--Lagrange equation for $ \tilde{L}=L^{\alpha} $ would conserve $L=L_{1}$ since $ \tilde{h}=(\alpha-1)L^{\alpha} $. All these solutions are solutions of the Euler--Lagrange equation for $L$ as well, thus $ L^{\alpha}\subset L $. In general, conservation of $ L_{1} $ is not guaranteed since $ L_{1}\rightarrow L_{1}+d\Lambda /d\tau $ is also a
homogeneous Lagrangian of order one equivalent to $ L_{1} $. This suggests
that there may be a choice of $ \Lambda $, a ``gauge fixing'', such that $L_{1}+d\Lambda /d\tau $ is conserved even if $ L_{1} $ is not. The above
discussion applies to a more general homogeneous Lagrangians as well.\cite{Fairlie 1999}

In the example of the relativistic particle, the Lagrangian and the trajectory parameterization have a geometrical meaning. In general, however, parameterization of a trajectory is quite arbitrary for any observer. If there is no preferred trajectory parameterization in a smooth space-time, then we are free to choose the standard of distance (time, using natural units $c=1$). Thus, \textit{our theory should not depend on the choice of parameterization}. By inspection of the Euler--Lagrange equations, any homogeneous Lagrangian of order $n$ ($L(x,\alpha \vec{v})=\alpha ^{n}L(x,\vec{v}) $) provides a reparametrization invariant equations ($ \tau\rightarrow \tau /\alpha,\vec{v}\rightarrow \alpha \vec{v}$).  
Next, note that the action $ S $ involves an integration that is a natural structure for orientable manifolds ($ M $) with an $ n $-form of the volume. Since a trajectory is
a one-dimensional object, then what we are looking at is an embedding $
\phi:\Bbb{R}^{1}\rightarrow M$. This means that we push forward the
tangential space $ \phi _{*}:T(\Bbb{R}^{1})=\Bbb{R}^{1}\rightarrow T(M)$,
and pull back the cotangent space $ \phi ^{*}:T(\Bbb{R}^{1})=\Bbb{R}
^{1}\leftarrow T^{*}(M)$. Thus a 1-form $ \omega $ on $ M $ that is in $
T^{*}(M) $ ($ \omega =A_{\mu}\left(x\right) dx^{\mu} $) will be pulled
back on $ \Bbb{R}^{1} $ ($ \phi ^{*}(\omega) $) and there it should be
proportional to the volume form on $ \Bbb{R}^{1} $ ($ \phi
^{*}(\omega)=A_{\mu}\left(x\right) (dx^{\mu}/d\tau)d\tau \sim d\tau $),
allowing us to integrate $ \int \phi ^{*}(\omega) $ : 
\[
\int \phi ^{*}(\omega)=\int Ld\tau =\int A_{\mu}\left(x\right)
v^{\mu}d\tau. 
\]
Therefore, by selecting a 1-form $ \omega =A_{\mu}\left(x\right) dx^{\mu}
$ on $ M $ and using $ L=A_{\mu}\left(x\right) v^{\mu} $ we are actually
solving for the embedding $ \phi:\Bbb{R}^{1}\rightarrow M $ using a chart
on $ M $ with coordinates $ x:M\rightarrow \Bbb{R}^{n}$. The Lagrangian
obtained this way is homogeneous of first order in $ v $ with a very
simple dynamics. The corresponding Euler--Lagrange equation is $ F_{\nu\mu}v^{\mu}=0 $ where $ F $ is a 2-form ($ F=dA $) -- the Faraday's tensor. If  the assumption that $ L $ is a pulled back 1-form is relaxed and instead one assumes that it is just a homogeneous Lagrangian of order one, then the corresponding reparametrization-invariant theory may have an interesting dynamics.

\textbf{First order homogeneous Lagrangians -- canonical form.}
Now we define what we mean by the \textit{canonical form of the first order homogeneous Lagrangian} and why do we prefer this mathematical expression. Let $S_{\alpha _{1}\alpha_{2}...\alpha _{n}}$ be a symmetric tensor of rank $n$  
which defines a homogeneous function of order $n$ ($S_{n}(\vec{v},...,\vec{v})=
S_{\alpha _{1}\alpha_{2}...\alpha_{n}}v^{\alpha_{1}}... v^{\alpha _{n}} $). 
The symmetric tensor of rank two is denoted by $ g_{\alpha \beta} $. Using this notation, the canonical form of the first order homogeneous Lagrangian is defined as: 
\begin{eqnarray}
\label{canonical form} 
L\left(\vec{x},\vec{v}\right)=\sum_{n=1}^{\infty}
\sqrt[n]{S_{n}\left(\vec{v},...,\vec{v}\right)}=A_{\alpha}v^{\alpha}+
\sqrt{g_{\alpha\beta}v^{\alpha}v^{\beta}}+...\sqrt[m]{S_{m}\left(\vec{v},...,
\vec{v}\right)}.
\end{eqnarray}

Any Lagrangian for the matter should involve interaction fields
that couple with the velocity $ \vec{v} $ to a scalar. When the matter action is
combined with the action for the interaction fields ($\mathcal S= \int
\mathcal{L} dV$), one obtains a full \textit{background independent theory}. Then the corresponding Euler--Lagrange equations contain ``dynamical derivatives'' on the left
hand side and sources on the right hand side: 
\[
\partial _{\gamma}\left(\frac{\delta \mathcal{L}}{\delta (\partial
_{\gamma}\Psi ^{\alpha})}\right) =\frac{\delta \mathcal{L}}{\delta \Psi
^{\alpha}}+\frac{\partial L_{matter}}{\partial \Psi^{\alpha}}. 
\]

The advantage of the canonical form of the first order homogeneous
Lagrangian (\ref{canonical form}) is that each interaction field, which is
associated with a symmetric tensor, has a unique matter source that is a
monomial in the velocities: 
\begin{equation}
\frac{\partial L}{\partial S_{\alpha _{1}\alpha _{2}...\alpha
_{n}}}=\frac{1}{n}
\left(S_{n}(\vec{v},...,\vec{v})\right) ^{\frac{1-n}{n}}v^{\alpha
_{1}}... v^{\alpha _{n}}. \label{sources}
\end{equation}

There are many other ways one can write first-order homogeneous 
functions.\cite{Rund 1966} For example, one can consider the following expression 
$L\left(\vec{x},\vec{v}\right) =\left(h_{\alpha
\beta}v^{\alpha}v^{\beta}\right) 
\left(g_{\alpha \beta}v^{\alpha}v^{\beta}\right) ^{-1/2} $. However, each of the
fields $h$ and $g$ has the same source type ($ \sim v^{\alpha}v^{\beta} $). At this stage, however, it is not clear why the same source type should produce different fields. Therefore, the canonical form (\ref{canonical form}) seems more appropriate for the current discussion.

\textbf{Extended objects.} 
In the previous sections, the classical mechanics of a point-like particle have been discussed as a problem concerned with the embedding 
$\phi:\Bbb{R} ^{1}\rightarrow M$. The map $\phi$ provides the trajectory
(the word line) of the particle in the target space $M$. In this sense,
we are dealing with a one dimensional object, the world-line of the particle (one dimensional W-brane). We think of an extended object as a manifold $W$ with dimension denoted by $D$. In this sense, we have to solve for $\phi:W\rightarrow M$ such that some action integral is minimized. From this point of view, we are dealing with mechanics of a brane. In other words, how is this $D$-dimensional extended object submerged in $ M$, and what are the relevant interaction fields? Following the
relativistic point particle discussion, we consider the space of the 
$D$-forms over the manifold $ M $, denoted by $\Lambda^{D}\left(M\right)$, that has dimension 
$\binom{m}{D}=\frac{m!}{D!(m-D)!}$. An element 
$\Omega $ in $ \Lambda ^{D}\left(M\right) $ has the form $ \Omega =\Omega
_{\alpha _{1}...\alpha _{m}}dx^{\alpha _{1}}\wedge dx^{\alpha _{2}}
\wedge... dx^{\alpha _{m}}$. We use the label $ \Gamma $ to index
different $ D $-forms over $ M, \Gamma =1,2,..., \binom{m}{D}$; thus
$\Omega
\rightarrow \Omega ^{\Gamma}=\Omega _{\alpha _{1}\dots\alpha _{m}}^{\Gamma}
dx^{\alpha _{1}}\wedge dx^{\alpha _{2}}\wedge... dx^{\alpha _{m}}$. Next we
introduce ``\textit{generalized velocity vectors}'' with components $
\omega ^{\Gamma} $ : 
\begin{eqnarray}
\omega ^{\Gamma}=\frac{\Omega ^{\Gamma}}{dz}=\Omega _{\alpha
_{1}...\alpha _{D}}^{\Gamma}\frac{\partial \left(x^{\alpha _{1}}x^{\alpha
_{2}}... x^{\alpha _{D}}\right)}{\partial (z^{1}z^{2}... z^{D})}, \quad dz=dz^{1}\wedge dz^{2}\wedge...\wedge dz^{D}.\nonumber
\end{eqnarray} In the above expression, $ \frac{\partial \left(x^{\alpha
_{1}}x^{\alpha _{2}}... x^{\alpha _{D}}\right)}{\partial
(z^{1}z^{2}... z^{D})} $ represents the Jacobian of the transformation from
coordinates $ \{x^{\alpha}\} $ over the manifold $ M $ to coordinates $
\{z^{a}\} $ over the brane. The pull back of a $ D $-form $ \Omega
^{\Gamma} $ must be proportional to the volume form over the brane: 
\begin{eqnarray}
\phi ^{*}\left(\Omega ^{\Gamma}\right)=\omega ^{\Gamma}dz^{1}\wedge...
\wedge dz^{D} =\Omega _{\alpha _{1}...\alpha
_{D}}^{\Gamma} \frac{\partial \left(x^{\alpha _{1}}... x^{\alpha
_{D}}\right)}{\partial (z^{1}... z^{D})}dz^{1}\wedge...
\wedge dz^{D}. \nonumber
\end{eqnarray} Thus, it is suitable for integration over the
$W$-manifold and the action is: 
\[ S\left[ \phi \right] =\int_{W}L\left(\vec{\phi},\vec{\omega}\right)
dz=\int_{W}\phi ^{*}\left(\Omega \right) =\int_{W}A_{\Gamma}
(\vec{\phi})\omega ^{\Gamma}dz. 
\] 
This is a homogeneous function in $ \omega $ and is reparametrization
(diffeomorphism) invariant with respect to the diffeomorphisms of the 
$W$-manifold. If we relax the linearity $L(\vec{\phi},\vec{\omega}) =\phi
^{*}\left(\Omega  \right) = A_{\Gamma}(\vec{\phi})\omega ^{\Gamma} $ in $ \vec{\omega}$,
then the canonical expression for the first order homogeneous Lagrangian is: 
\begin{eqnarray}\label{canonical d-brane L}
L\left(\vec{\phi},\vec{\omega}\right)=\sum_{n=1}^{\infty}\sqrt[n]{S_{n}
\left(\vec{\omega},...,\vec{\omega}\right)}=
A_{\Gamma}\omega^{\Gamma}+\sqrt{g_{\Gamma _{1}\Gamma _{2}}\omega ^{\Gamma
_{1}}\omega^{\Gamma_{2}}}+...
\end{eqnarray}

At this point, there is a strong analogy between the relativistic point
particle and the extended object. Some specific examples of $W$-brane theories correspond to the following familiar Lagrangians:

\vspace{.5cm}
\noindent
\textit{Lagrangian for a relativistic point particle} in an
electromagnetic field:  $\dim W=1$ (World-line) with $ \omega ^{\Gamma}\rightarrow
v^{\alpha}=\frac{dx^{\alpha}}{d\tau} $ 
\begin{eqnarray} L\left(\vec{\phi},\vec{\omega}\right)=A_{\Gamma}\omega
^{\Gamma}+ \sqrt{g_{\Gamma _{1}\Gamma _{2}}\omega ^{\Gamma _{1}}\omega ^{\Gamma _{2}}} \rightarrow qA_{\alpha}v^{\alpha}+
m\sqrt{g_{\alpha\beta}v^{\alpha}v^{\beta}}. \nonumber
\end{eqnarray}
\textit{Lagrangian for strings}: $\dim W=2$ (World-sheet) \\
$L\left(x^{\alpha},\partial _{i}x^{\beta}\right) =\sqrt{Y^{\alpha 
\beta}Y_{\alpha \beta}},$ with the following notation:
\begin{eqnarray}
\omega ^{\Gamma}\rightarrow Y^{\alpha \beta}=\frac{\partial (x^{\alpha},
x^{\beta})}{\partial (\tau,\sigma)}=\det \left(\begin{array}{cc}
\partial _{\tau}x^{\alpha} & \partial _{\sigma}x^{\alpha} \\ 
\partial _{\tau}x^{\beta} & \partial _{\sigma}x^{\beta}
\end{array}\right) =\partial _{\tau}x^{\alpha}\partial
_{\sigma}x^{\beta}-\partial _{\sigma}x^{\alpha}\partial
_{\tau}x^{\beta}.\nonumber
\end{eqnarray}
\textit{Dirac--Nambu--Goto Lagrangian (DNG)\cite{Pavsic 2001}}: 
$ L\left(x^{\alpha},\partial _{W}x^{\beta}\right)
=\sqrt{Y^{\Gamma}Y_{\Gamma}}. $
\vspace{.5cm}

The corresponding electromagnetic interaction term for W-banes is know as
Wess--Zumino term\cite{Bozhilov 2002} in string theory.

From the expressions (\ref{canonical form}) and (\ref{canonical
d-brane L}), one can see that the corresponding matter Lagrangians $(L)$, in
their canonical form, contain electromagnetic ($A$) and gravitational ($g$)
interactions, as well as interactions that are not clearly identified yet ($
S_{n},n>2$), if present at all in nature. At this stage, we have a theory with background fields since we don't know the equations for the interaction fields $A,$ $g,$ and $S_{n}$. To complete the theory, we need to introduce actions for these interaction
fields. If one is going to study the new interaction fields $S_{n},n>2$,
then some guiding principles for writing field Lagrangians are needed.

One approach is to apply the above discussion and view the $S_{n}$ fields as related to an embedding of the M-manifold into the manifold of symmetric tensors over M. Another approach would be to use the external derivative $d$, external multiplication $\wedge $
, and Hodge dual $*$ operations in the external algebra $\Lambda \left(
T^{*}M\right) $ over $M$ to construct objects proportional to the volume
form over $M$. For example, for any $n$-form $(A)$ the expressions $A\wedge
*A$ and $dA\wedge *dA$ are forms proportional to the volume form. 
The next important principle comes from the symmetry in the matter equation.
That is, if there is a transformation $A\rightarrow A^{\prime }$ that leaves
the matter equations unchanged, then there is no way to distinguish $A$ and $
A^{\prime }$. Thus the action for the field $A$ should obey the same gauge
symmetry. For the electromagnetic field ($A\rightarrow A^{\prime }=A+\mathrm{
d}f$) this leads uniquely to the field Lagrangian $\mathcal{L}=dA\wedge *dA=F\wedge
*F $, when for gravity\cite{VGG Kiten 2002} it leads to the Cartan--Einstein action\cite{Adak et al 2001} $S\left[R\right] =\int R_{\alpha \beta }\wedge *(\mathrm{d}x^{\alpha }\wedge dx^{\beta })$. 

\textbf{Non-relativistic limit.}
For a $W$-brane we assume the existence a local coordinate frame where one component of the generalized velocity can be set to 1 ($\omega ^{0}=1$). This generalized velocity component is associated with the brane ``time coordinate.'' In fact, $\omega ^{0}=1$ means that there is an integral embedding of the brane in the target space $M$, and the image of the brane is a sub-manifold of $M$. If the coordinates of $M$ are labeled so that $x^{i}=z^{i},i=1,...,D$, then $x^{i}$ are internal coordinates that can collapse to only one coordinate -- the ``world line''. This provides a gauge-fixing that allows one to do canonical quantization. This approach is mainly concerned with the choice of a coordinate time that is used as the trajectory parameter.\cite{Nikitin-string theory,Dirac 1958,Henneaux and Teitelboim,Schwinger 1963} Such choice removes the reparametrization invariance of the theory.

In a local coordinate system where $\omega ^{0}=1$ and the metric is a
``one-time-metric'' we have: 
\begin{eqnarray*}
L &=&A_{\Gamma }\omega ^{\Gamma }+\sqrt{g_{\Gamma _{1}\Gamma _{2}}\omega
^{\Gamma _{1}}\omega ^{\Gamma _{2}}}+...+\sqrt[m]{S_{m}\left( \vec{\omega}
,...,\vec{\omega}\right) }\rightarrow \\
&\rightarrow &A_{0}+A_{i}\omega ^{i}+\sqrt{1-g_{ii}\omega ^{i}\omega ^{i}}
+...\approx A_{0}+A_{i}\omega ^{i}+1-\frac{1}{2}g_{ii}\omega ^{i}\omega
^{i}+....
\end{eqnarray*}
Thus the Hamiltonian function is not zero anymore, so we can do canonical
quantization, and the Hilbert space consists of the functions $\Psi \left(
x\right) \rightarrow \Psi \left( z,\tilde{x}\right) $ where $\tilde{x}
=x^{i},i=D+1,...,m$. The brane coordinates $z$ should be treated as $t$ in
quantum mechanics in the sense that the scalar product should be an integral
over the space coordinates $\tilde{x}$. For $W$-branes the one-time coordinate reflects separation of the internal from the external coordinates when the $W$-brane is considered as a sub-manifold of the target space manifold $M$.

Even though canonical quantization can be applied after the above gauge fixing, one is not usually happy because the covariance of the theory is lost and time is a privileged coordinate. In general, there are well developed procedures for covariant 
quantization.\cite{Nikitin-string theory,Henneaux and Teitelboim,Dirac 1958a,Teitelboim 1982,Sundermeyer 1982} In this paper, however, we are not going to discuss these methods. Instead, we will employ a different quantization strategy\cite{VGG Varna 2002}, but before that we will discuss the mass-shell and Klein--Gordon equations.

\textbf{The mass-shell and Klein--Gordon equation.}
Since the functional form of the canonical Lagrangian is the same for any $W$-brane, we use $v$, but it could be $\omega $ as well. We define the momentum $p$ and generalized momentum $\pi $ for our canonical Lagrangian as follow: 
\begin{eqnarray*}
p_{\Gamma } &=&\frac{\delta L\left( \phi ,\omega \right) }{\delta \omega
^{\Gamma }}=eA_{\Gamma }+m\frac{g_{\Gamma \Sigma }\omega ^{\Sigma }}{\sqrt{
g\left( \vec{\omega},\vec{\omega}\right) }}+...+\frac{S_{\Gamma \Sigma
_{1}...\Sigma _{n}}\omega ^{\Sigma _{1}}...\omega ^{\Sigma _{n}}}{\left(
S\left( \omega,...,\omega \right) \right) ^{1-1/n}}+..., \\
\pi _{\alpha } &=&p_{\alpha }-eA_{\alpha }-...\frac{S_{\alpha \beta
_{1}...\beta _{n}}v^{\beta _{1}}... v^{\beta _{n}}}{\left( S\left(
v,...,v\right) \right) ^{n/\left( n+1\right) }}...=m\frac{g_{\alpha \beta
}v^{\beta }}{\sqrt{g\left( \vec{v},\vec{v}\right) }}.
\end{eqnarray*}
In the second equation we have used $v$ instead of $\omega $ for simplicity.
Notice that this generalized momentum ($\pi $) is consistent with the usual
quantum mechanical procedure $p\rightarrow p-eA$ that is used in Yang--Mills theories, as well as with the usual GR expression $p_{\alpha }=mg_{\alpha\beta }v^{\beta }$. Now it is easy to recognize the mass-shell constraint as a mathematical identity: 
\[
\frac{\vec{v}}{\sqrt{\vec{v}^{2}}}\cdot \frac{\vec{v}}{\sqrt{\vec{v}^{2}}}
=1\Rightarrow \pi _{\alpha }\pi ^{\alpha }=m^{2} \Rightarrow 
\left( \vec{p}-e\vec{A}-\vec{S}_{3}\left( v\right) -\vec{S}_{4}\left(
v\right) -...\right) ^{2}\Psi =m^{2}\Psi . 
\]
Notice that ``gravity'' as represented by the metric is gone, while the
Klein--Gordon equation appears. The $v$ dependence in the $S$ terms reminds
us about the problem related to the change of coordinates $(x,v)\rightarrow
(x,p)$. So, at this stage we may proceed with the Klein--Gordon equation, if we wish.

\textbf{Dirac equation from H=0.}
An interesting approach to the Dirac equation has been suggested by 
H. Rund.\cite{Rund 1966} The idea uses Hamiltonian linear in the momentum ($H=\gamma ^{\alpha}p_{\alpha }$) and the base manifold principle group $G$. To have the Hamiltonian $H$ invariant under $G$-transformations, the $\gamma$ objects should transform appropriately and provide also a realization of the generators of G. Since we want $\gamma $ and $p $ to transform as vectors, it is clear that $p$ should be a covariant derivative, but what is its structure? Consider a homogeneous Lagrangian that can be written as $L\left(\phi,\omega \right)
=\omega ^{\Gamma}p_{\Gamma}=\omega ^{\Gamma}\partial L\left(\phi,\omega
\right) /\partial \omega ^{\Gamma} $ with a Hamiltonian function that is
identically zero: $h=\omega ^{\Gamma}\partial L\left(\phi,\omega \right)
/\partial \omega ^{\Gamma}-L\left(\phi,\omega \right) \equiv 0$. Notice that 
$\omega ^{\Gamma} $ is the determinant of a matrix (the Jacobian of a
transformation\cite{Fairlie and Ueno}); thus $\omega ^{\Gamma}\rightarrow
\gamma ^{\Gamma} $ seems an interesting option for quantization. Even more,
for the Dirac theory we know that $\gamma ^{\alpha} $ are the `velocities' ($
dx/d\tau =\partial H/\partial p $).

If we quantize using ($h\rightarrow H$), then the space of physical states
should satisfy: $H\Psi =0$. By applying $\omega ^{\Gamma }\rightarrow \gamma
^{\Gamma }$, which means that the (generalized) velocity is considered as a
vector with non-commutative components, we have $\left( \gamma ^{\Gamma
}p_{\Gamma }-L\left( \phi ,\gamma \right) \right) \Psi =0$. For a point particle,
using the canonical form of the Lagrangian (\ref{canonical form}) and the
algebra of the $\gamma $ matrices following Run's approach\cite{Rund 1966} this gives: 
\begin{eqnarray*}
H &=&\gamma ^{\alpha }p_{\alpha }-L\left( \phi ,\gamma \right) =\gamma
^{\alpha }p_{\alpha }-eA_{\alpha }\gamma ^{\alpha }-m\sqrt{g_{\alpha \beta
}\gamma ^{\alpha }\gamma ^{\beta }}-...\sqrt[m]{S_{m}\left( \vec{\gamma},...,
\vec{\gamma}\right) }, \\
&\rightarrow &\gamma ^{\alpha }p_{\alpha }-eA_{\alpha }\gamma ^{\alpha
}-m-...\sqrt[2m]{S_{2m}g^{m}}-...\sqrt[2n+1]{S_{2n+1}g^{n}\gamma }...
\end{eqnarray*}
Since $g_{\alpha \beta }$ is a symmetric tensor such that $\{\gamma
^{\alpha },\gamma ^{\beta }\}\sim g^{\alpha \beta }$, then $g_{\alpha \beta
}\gamma ^{\alpha }\gamma ^{\beta }\sim g_{\alpha \beta }\{\gamma ^{\alpha
},\gamma ^{\beta }\}\sim g_{\alpha \beta }g^{\alpha \beta }\sim 1$.
Therefore, gravity seems to leave the picture again. The symmetric structure
of the extra terms $S_{m}$ can be used to reintroduce $g$ and to reduce the
powers of $\gamma $. Thus the higher even order terms contribute to the mass $m$,
making it variable\cite{Bekenstein 1993} with $\vec{x}$.

\textbf{Summary.} We have discussed the structure of the matter Lagrangian for extended objects. Imposing reparametrization invariance of the action $S$
naturally leads to a first order homogeneous Lagrangian. In its canonical
form, $L$ contains electromagnetic and gravitational interactions, as well
as interactions that are not yet identified. The non-relativistic limit for a brane has been defined as those coordinates where the brane is an integral sub-manifold of the target space. This gauge can be used to remove reparametrization invariance of the action $S$ and make the Hamiltonian function suitable for canonical quantization.
The existence of a mass-shell constraint is universal. It is essentially due
to the gravitational (quadratic in velocities) type interaction in the
Lagrangian and always leads to a Klein--Gordon like equation. Once the algebraic properties of the $\gamma$-matrices are defined, one can use $v\rightarrow \gamma $ quantization in the Hamiltonian function $h=pv-L\left( x,v\right) $ to obtain the Dirac equation.

\end{document}